\documentclass[reprint,twocolumns,amsmath,amssymb,aps]{revtex4-2}
\usepackage{graphicx}
\usepackage{dcolumn}
\usepackage{bm}
\usepackage{url}
\usepackage[colorlinks=true,urlcolor=blue,linkcolor=blue,citecolor=blue
]{hyperref}
\usepackage{placeins} 
\usepackage{xcolor}
\usepackage{xparse}
\usepackage[T1]{fontenc}
\usepackage[latin9]{inputenc}

\NewDocumentCommand{\tens}{t_}
 {%
  \IfBooleanTF{#1}
   {\tensop}
   {\otimes}%
 }
\NewDocumentCommand{\Log}{o}{%
  \IfNoValueTF{#1}{}{{}^{#1}\!}\log}%

\newcommand{\ket}[1]{| #1 \rangle}

\begin{document}


\title{Negative correlations can play a positive role in  disordered quantum walks}

\author{Marcelo A. Pires}
\email{piresma@cbpf.br}
\affiliation{Centro Brasileiro de Pesquisas F\'isicas, Rua Dr Xavier Sigaud 150, 22290-180 Urca, Rio de Janeiro-RJ, Brazil}

\author{S\'\i lvio M. \surname{Duarte~Queir\'{o}s}}
\altaffiliation[Associate to the ]{National Institute of Science and Technology for Complex Systems, Brazil}
\email{sdqueiro@cbpf.br}
\affiliation{Centro Brasileiro de Pesquisas F\'isicas, Rua Dr Xavier Sigaud 150, 22290-180 Urca, Rio de Janeiro-RJ, Brazil}

\date{\today}

\begin{abstract} 
We investigate the emerging properties of quantum walks with temporal disorder engineered from a binary Markov chain with tailored correlation, $C$, and disorder strength, $r$. We show that when the disorder is weak --- $r \ll 1$ --- the introduction of negative correlation leads to a counter-intuitive higher production of spin-lattice entanglement entropy,  $S_e$, than the setting with positive correlation, that is $S_e(-|C|)>S_e(|C|)$. These results show that negatively correlated disorder plays a more important role in quantum entanglement than it has been assumed in the literature.
\end{abstract}

\maketitle


\section{\label{sec:intro}Introduction}

Inasmuch as the random walk has been at the cradle of the development of processes and techniques through out one hundred-off years, the introduction of its quantum counterpart, the quantum walk~\cite{aharonov1993quantum}(QW), urged a range of prospective applications, namely those related to the Feynman's quantum computer proposal made some ten years earlier~\cite{feynman1982simulating}.
Formally defined by a succession of local and unitary operations on qubits, QWs have definitely established as the direct path to understand complex quantum phenomena by means of relatively simple protocols~ \cite{ambainis2003quantum,venegas2008quantum,portugal2013quantum} that can be reproduced in a laboratory\cite{wang2013physical,grafe2016integrated,neves2018photonic} or the development of quantum algorithms~\cite{portugal2013quantum}.
\textcolor{black}{
Explicitly, the quantum walk evolves on a Hilbert space, $\mathcal{H}_2\otimes \mathcal{H}_\mathbb{Z}$, by means of the combined application of two unitary operators
, the operator $\hat{R}$ acts on subspace $\mathcal{H}_2$ and plays the role of quantum coin related to internal (spin) states, $s$, whereas the external states related to the subspace $\mathcal{H}_\mathbb{Z}$ change due to the \emph{shift operator}, $\hat{T}$.
%
%
%
%
The successive application of a time-evolution operator  
%
%
%
%
allows obtaining the  probability of finding the walker at position $x$ at time $t$, $P_t(x)$, is given by
\mbox{$P_t(x)= \sum _{s=\{D,U\}}  \left| \Psi ^{\left(s\right)}  _t \left( x \right) \right|^2$}, from which the statistical characterization of the QW is made.
}
Along the years, the original model~\cite{aharonov1993quantum} has given raise to multitude of variants, e.g., by changing the dimensionality of the walk, the topology of the lattice as well as disorder in the jump size and the angle defining the quantum coin~\cite{ribeiro2004aperiodic,banuls2006quantum,kendon2006random,attal2012open,uchiyama2018environmental,zeng2017discrete,vieira2013dynamically,di2018elephant,pires2019multiple,pires2020quantum,vakulchyk2019wave,pires2020genuine}.

A comparison between the classical and the quantum walker protocol shows that in latter, the stochastic part is replaced by operations on an internal degree of freedom the walker --- traditionally its spin. Given that the position state of the walker has to do with that internal state, it is therefore natural to ask how much both states relate. The most  straightforward way to quantum mechanically assess the degree of relation between both states is to look at quantum entanglement, specifically the entanglement entropy. 
\textcolor{black}{
The study of entanglement is of crucial importance for quantum information theory either for fundamental or applied issues \cite{Horodecki2009}. }
In this work, we analyse the impact in the quantum entanglement of the correlation function --- a classical measure --- defining the disorder introduced in the system by means of a Markov process for the coin operator angle. Those protocols allow experimental implementation with the state-of-art photonic platforms.~\cite{wang2018dynamic}Especially, we focus on the relevance of negative correlations to enhancing or undermining entanglement.

As we will show,  the arrangement of random disorder with negative correlation leads to the appearance of distinctive configurations that produce more spin-lattice entanglement than the corresponding case with positive correlation.


\begin{figure*}[t]
\centering
\includegraphics[scale=0.50]{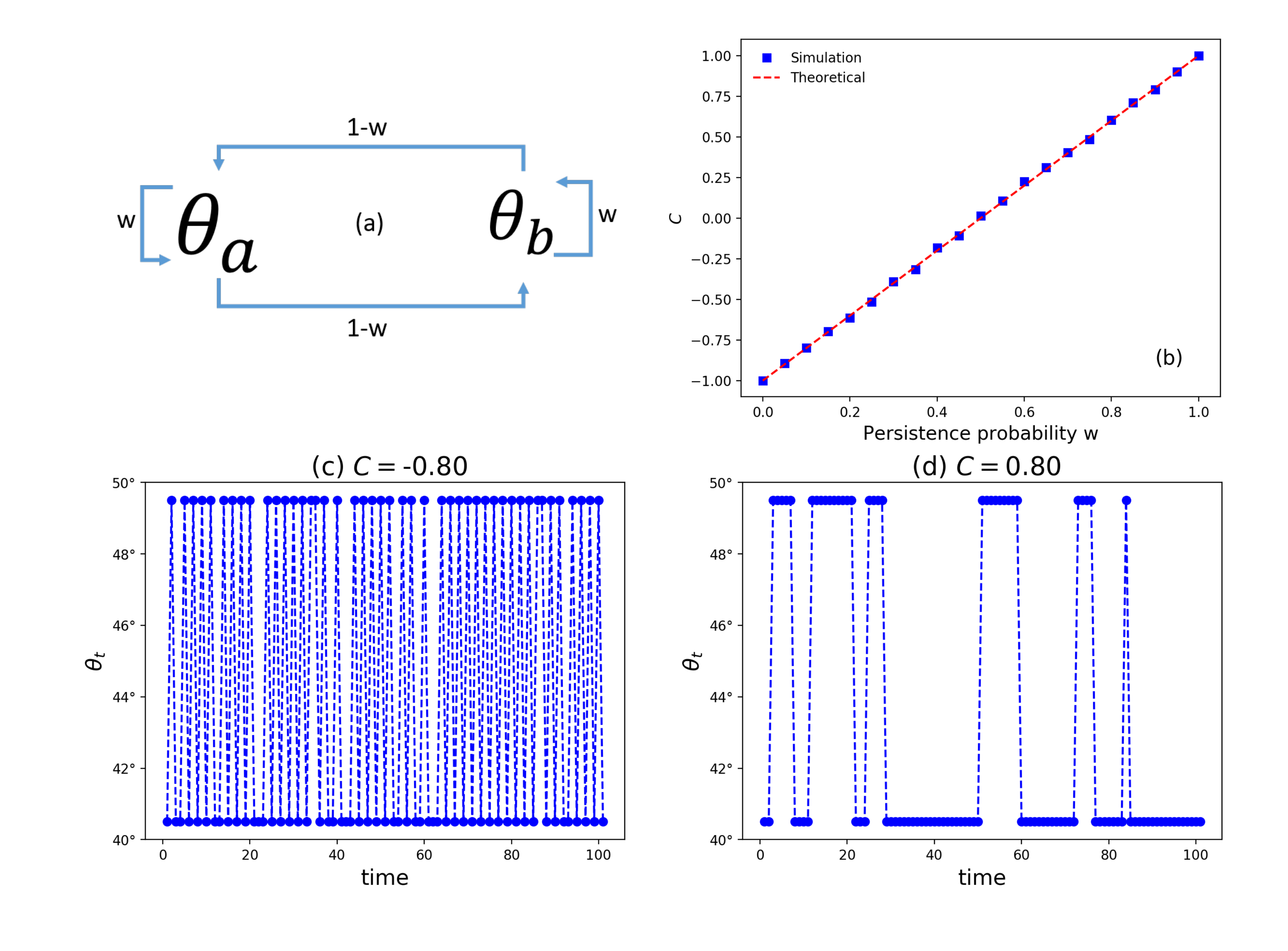}
\caption{(a) Markov chain employed for tailoring the disorder with prescribed   persistence probability $w$ that provides the control of the autocorrelation with Eq.\ref{eq:corr_w}. (b) Autocorrelation of the sequence 
$\{\theta_t\}_{t=0}^{T}$. Monte Carlo simulations performed until $t_{max}=5000$. The theoretical line comes from Eq.\ref{eq:corr_w}. (c-d) Time series for $\theta_t$ engineered from  Markov  chains with $\theta_a=(1+r)\pi/4$ and 
$\theta_b=(1-r)\pi/4$ for $r=0.1$ and $C=\pm 0.8$.} 
\label{Fig:theta_t}
\end{figure*}
%

\section{\label{sec:rev-lit}Literature review }

Carrying out computational simulations, the issue of entanglement between the internal and external degrees of freedom of a quantum walker was first studied in 
2005;~\cite{carneiro2005entanglement} there, it was conveyed the asymptotic coin-position entanglement entropy
$S_{\rm e} \rightarrow 0.872\ldots $. Afterwards, that result 
was analytically corroborated~\cite{abal2006quantum} and 
verified in linear-optical experiments.~\cite{su2019experimental}

Those works only considered disorder-free configurations in which the coin and translation operator are homogeneous in space and time. However, as mentioned in the previous Section, the assumption of disorder in such operators has opened the door for  novel phenomenologies.  For instance, nongaussian distributions can emerge~\cite{shikano2014discrete} or unusual gaussian with hyperballistic spreading can take place \cite{di2018elephant}.
In 2012, it was reported an enhancement of entanglement with random disorder~\cite{chandrashekar2012disorder}, a result that was analytically proven afterwards~\cite{vieira2013dynamically}. 
Subsequently, the entanglement analysis was extended for other protocols of disorder either in the coin operator~\cite{chandrashekar2012disorder,vieira2013dynamically,salimi2012asymptotic,rohde2013quantum,vieira2014entangling,di2016discrete,orthey2019weak,singh2019accelerated,montero2016classical,vieira2014entangling,zeng2017discrete,buarque2019aperiodic,wang2018dynamic} or in the step operator~\cite{sen2019scaling,mukhopadhyay2020persistent,pires2020quantum}. 
Recently, it was brought forth the first model with disorder in the translation operator that displays both the strengthening of entanglement and tunable spreading from slower-than-ballistic to faster-than ballistic\cite{pires2019multiple}. From an experimental perspective, using photonic platforms, it was possible to confirm non-static disorder can favor entanglement~\cite{wang2018dynamic}.

QWs with correlated disorder were previously studied in both  discrete-\cite{romanelli2007sub,ahlbrecht2011asymptotic,mendes2019localization} and continuous-time\cite{yin2008quantum,rossi2017continuous} representations. In Refs.~\cite{ahlbrecht2011asymptotic,yin2008quantum}, it was shown that randomness leads to a diffusive-like behavior even when correlation is present. Regardless, the question of how the coin-space entanglement is affected by the correlation in the disorder remains uncovered to the best of our knowledge.

\begin{figure}[t]
\centering
\includegraphics[scale=0.58]{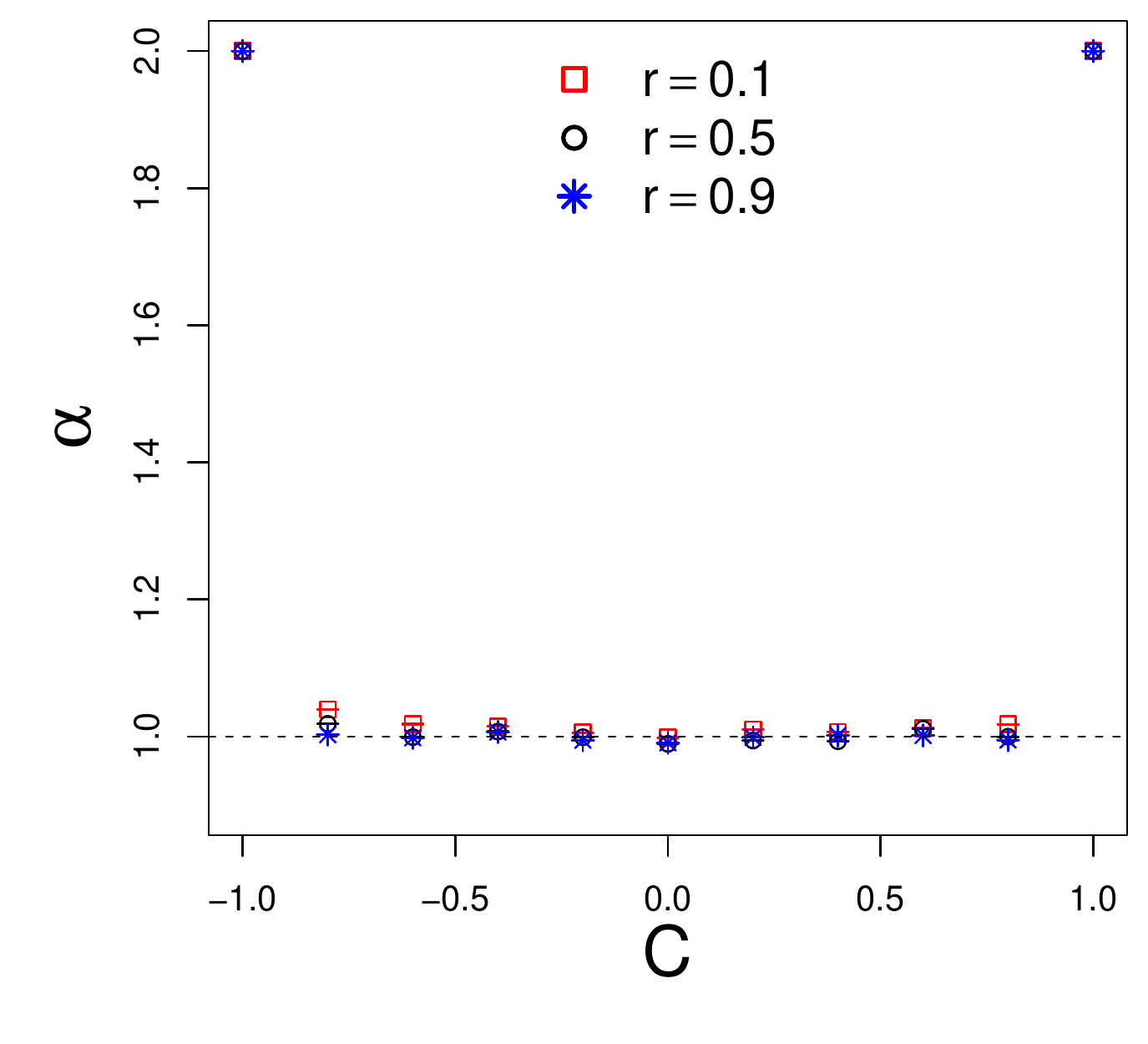}
\caption{Diagram of the spreading regime versus the correlation $C$ in the disorder for typical values  $r=\{01,0.5,0.9\}$. We compute $\alpha$ from $m_2(t)\sim t^\alpha$ employing $t_{max}=5\times10^5$ after discarding the transient $t<5\times10^4$.}
\label{Fig:alpha-vs-cor}
\end{figure}

\section{\label{sec:model}Model}

\subsection{One-dimensional quantum walks}

We consider a discrete-time evolution of a two-state quantum walk moving on a one-dimension lattice whose composite Hilbert space is 
$\mathcal{H}_2\otimes \mathcal{H}_\mathbb{Z}$. 
The composed state $|x\rangle\otimes |c\rangle =|x,c \rangle$  indicates the position  $x \in \mathbb{Z}$ of a QW with  internal degree of freedom (up/down),  $ c=\{U,D\}$. Such  degree of freedom,  $ c=\{U,D\}$, is associated with the corresponding space-time dependent amplitude of probability  $\psi_t^{U,D}(x) $, respectively.  
 At a given time 
 $t \in \mathbb{N}$ we can write the full wave function
 $\Psi_t$ as
\begin{align}
\Psi_t 
=
\sum_{x \in \mathbb{Z}} 
\left(
\psi_t^U(x) 
\ket{U}
+
\psi_t^D(x) 
\ket{D}
\right)
\otimes
\ket{x} 
\label{Eq:psi_geral}
\end{align}

The time evolution of the QW is governed by 
\begin{align}
\Psi_{t+1}=\hat W \Psi_t
\label{Eq:WSC1}
\end{align}
\begin{align}
\hat W = \hat{T}(\hat{R}\otimes \mathcal{I}_\mathbb{Z})
\label{Eq:WSC2}
\end{align}

with the identity operator $\mathcal{I}_\mathbb{Z}=\sum_{x \in \mathbb{Z}} |x \rangle \langle x|$ and: 
\begin{itemize}
    \item The coin operator:
  \begin{align}
    \widehat{R}: 
    \begin{cases}
     |x, U \rangle \rightarrow 
     \cos\theta_t |x, U \rangle +
     \sin\theta_t |x, D \rangle
      \\
     |x, D \rangle \rightarrow 
     \sin\theta_t |x, U \rangle 
     -\cos\theta_t |x, D \rangle
    \end{cases}
    \label{Eq:C-explicito}
  \end{align}
   where the off-diagonal elements  modulated by $\sin\theta_t$ are accountable for the coupling between the evolution of $\psi_t^U(x)$ and $\psi_t^D(x) $. The diagonal elements tempered by $\pm\cos\theta_t$ are responsible for propagation. The time-dependence of $\theta_t$ will be described in detail below.
    \item  The state-dependent translation operator: 
  \begin{align}
    \widehat{T}: 
    \begin{cases}
     |x, U \rangle \rightarrow |x+1, U \rangle 
      \\
     |x, D \rangle \rightarrow |x-1, D \rangle 
    \end{cases}
    \label{Eq:T-explicito}
  \end{align}
Meaning that the hopping-induced flux of probability  takes place to neighbor sites.
\end{itemize}

Now we set the initial condition as
\begin{equation}
 \psi_0^U(x) = \frac{1}{\sqrt{2}} \delta_{x,0}  
,\quad  
\psi_0^D(x) = \frac{ i }{\sqrt{2}} \delta_{x,0} 
,  
 \label{Eqt0}
 \end{equation}

\subsection{\label{sec:theta_t} Tailoring disorder with Markov chains}

At each time step $t$, we generate a random variable $z_t = \{a=-1,b=1\}   $ following a 
two-state Markov chain   with transition probability matrix: 
\begin{equation}
M
= 
\begin{pmatrix}
p_{a \rightarrow a}  &  p_{a \rightarrow b} \\
p_{b \rightarrow a} &  p_{b \rightarrow b}
\end{pmatrix}
\quad
=
\begin{pmatrix}
w &  1-w \\
1-w  & w
\end{pmatrix}
\label{eq:tr_mat}
\end{equation}
 where $w$ is the persistence probability. Explicitly, $w$ quantifies the probability of a given value to persist in the same state in the next step $t+1$. Thus, the switching probability is $1-w$. See Fig.\ref{Fig:theta_t}(a).

We set the baseline angle $\theta_0$, the kicking strength $r$. Then we define $\theta_a=(1+r)\theta_o$ and 
$\theta_b=(1-r)\theta_o$.  This protocol is complete with the equation
\begin{equation}
 \theta_t  = \frac{(1-z_t)\theta_a}{2}
   +  \frac{(1+z_t)\theta_b}{2}
=   
\begin{cases}
  \theta_a 
     &   
     \text{ if } z_t=-1
     \\
   \theta_b  
& 
\text{ if } z_t=1
   \end{cases}
\end{equation}
 
With this routine, each kick train has a changeable length, but constant amplitude $(1 \pm r)\theta_o$.
  In all circumstances, we fixed 
  $\theta_o=\pi/4$. As $0\leq r\leq 1$, then 
$\theta_a=(1+r)\pi/4\in [\pi/4,\pi/2]$ and 
$\theta_b=(1-r)\pi/4\in [0,\pi/4]$, 
then $0\leq (1-r)\pi/4\leq \theta_t\leq (1+r)\pi/4\leq \pi/2$.

For characterizing the similarity of the patterns through $\theta_t$ generated over time we employ the autocorrelation
\begin{equation}
C = 
\langle  z_t z_{t-1} \rangle = \frac{1}{T-1} \sum_{t=1}^{T} z_t z_{t-1}  = 2w - 1.
\label{eq:corr_w}
\end{equation}
In Fig.~\ref{Fig:theta_t}(b), we show the estimated autocorrelation obtained from the Monte Carlo simulation of the disorder and using $C=\langle  z_t z_{t-1} \rangle$.
 
Depending on the magnitude of the correlation we impose on the system The patterns in $\theta_t$ show different levels of similarity during the temporal evolution . 
The time series displayed in Fig.~\ref{Fig:theta_t}(c-d) illustrate the antipersistent patterns for $C=-0.8$ (marked by pulse trains of short duration) as well as the persistent patterns for $C=0.8$ (marked by pulse trains of long duration).
 
Apart from the transition matrix in Eq.~(\ref{eq:tr_mat}), we generate a sequence, $\{\theta_t\}_{t=0}^{T}$, with the prescribed correlation $C$, which has the interesting property of being unbiased since the fractions of each ingredient in $\{\theta_t\}_{t=0}^{T}$ are
\begin{equation}
f_a = \frac{p_{ab}}{p_{ba}+p_{ab}} = 50\% \quad t>>1,
\end{equation}
\begin{equation}
f_b = \frac{p_{ba}}{p_{ba}+p_{ab}} = 50\%  \quad t>>1,
\end{equation}
where $p_{ba}=p_{ab}=1-w$.

\section{\label{sec:results}Results and discussion}

We start the characterization of our dynamics by computing the scaling exponent $\alpha$ of $m_2(t)\sim t^\alpha$, the second statistical moment of the probability distribution $P_t(x)$. With $\alpha $, we classify this process in terms of its diffusion features. Explicitly,
\begin{equation}
\alpha = \lim_{t \to \infty} \frac{\log m_2(t)}{\log t}
\end{equation}
\begin{equation}
m_2(t) = \overline{x^2}_t =\sum_x x^2P_t(x) .
\end{equation}
\begin{equation}
P_t(x) = |\psi_{t}^{D}(x)|^2 + |\psi_{t}^{U}(x)|^2 
\end{equation}

\begin{figure*}[!hbtp]
\centering

\includegraphics[scale=0.59]{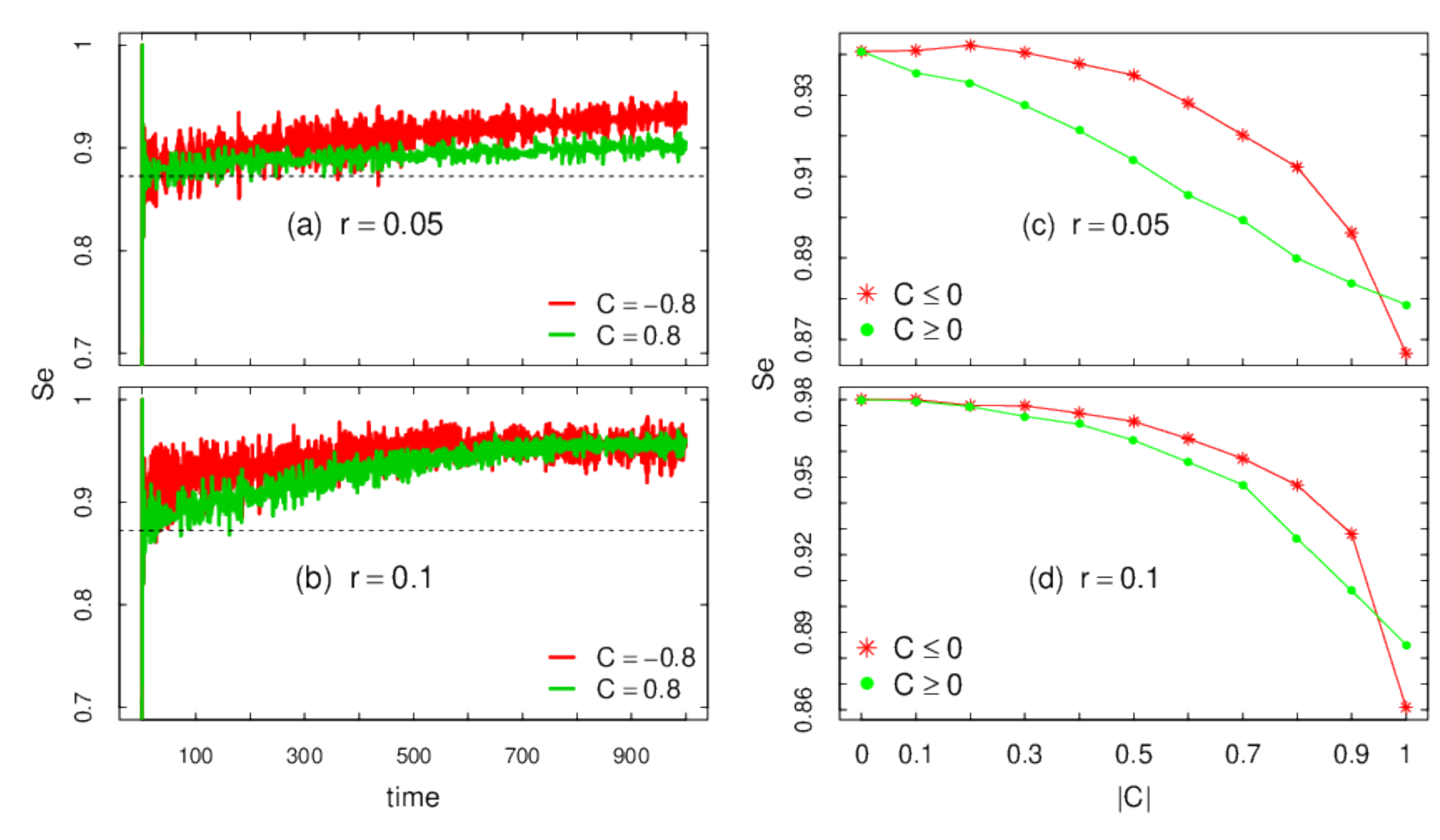}

\caption{Von Neumann entanglement entropy $S_e$ versus (a-b) time, (c-d) $C$. Other parameters displayed in the panels. For (c-d) $S_e$ is computed at $t=500$ and each point is an average over 100 samples.}
\label{fig:se-t-c-r}
\end{figure*}

In Fig.\ref{Fig:alpha-vs-cor}, we see how $\alpha$ depends on the correlation in the disorder $C$. Let us first consider the particular cases previously addressed in the literature.
For uncorrelated disorder $C=0$, the diffusive behavior $\alpha = 1$ is achieved as expected (see for instance Refs.~ \cite{wojcik2003diffusive,kovsik2006quantum,joye2011random,ahlbrecht2012asymptotic}). 
For $C=1$, $\theta_t$ is constant, which explains the ballistic spreading corresponding to $\alpha=2$. In the extreme anti-correlated case, $C=- 1$, $\theta_t$ is  periodically alternating~\cite{machida2010limit,rousseva2017alternating} and this deterministic pattern  leads to ballistic spreading as the standard quantum walk.
For the other configurations, we observe that randomness induces a diffusive-like scaling exponent $\alpha$ for any correlation $|C|<1$. Even though this is not straightforward to interpret for a strong correlation $0.7<|C|<1$ this result is in agreement with Refs.~\cite{ahlbrecht2011asymptotic,yin2008quantum}. 
That is to say, that randomness plays an important role in the path towards diffusive behavior in quantum walks with time-dependent coins but spatial translational invariance. 
Event though randomness is an important ingredient for the emergence of diffusive scaling in quantum walks; we emphasize it is not the only path to diffusion in quantum walks.
For instance, two counterexamples are presented in Refs.~\cite{romanelli2009driving,panahiyan2018controlling} where it was reported that some specific time dependence in $\theta_t$ can lead to diffusive-like spreading without any randomness.
In respect of spreading, the protocol in this manuscript shows its adequacy to recover the known phenomenology.

We now move to our main findings. The qubit-lattice entanglement is an important quantity that can also be quantified experimentally.  
To assess such feature we compute the von Neumann entropy
\begin{equation}
S_e  = - \mathrm{Tr} \left[ \rho^c\log\rho^c \right],
\end{equation}
where $\rho^c$ is the reduced density matrix obtained after tracing out the position degree of freedom,
\begin{equation}
\rho^c = \mathrm{Tr}_x(\rho) 
\end{equation}
and where $\rho$ is the full density matrix of the complete system
\begin{equation}
\rho = | \Psi \rangle \langle  \Psi | .
\end{equation}
Following ~\cite{abal2006quantum,vieira2014entangling,zeng2017discrete}, we can obtain $S_e$ with expressions that are computationally faster to process
\begin{equation}
 S_{\rm e}  = - \sum_{\lambda \in \{\lambda^{\pm}\}}
 \lambda \log_2 \lambda
\end{equation}
where the eigenvalues $\lambda^{\pm}$ of $\rho^c$ are
\begin{equation}
 \lambda^{\pm} = \frac{1}{2} \pm \sqrt{\frac{1}{4}-G_uG_d + |G_{ud}|^2 },
\end{equation}
with 
\begin{equation}
G_u = \sum_x |\psi_{t}^{U} (x)|^2    
\end{equation}
\begin{equation}
G_d = \sum_x |\psi_{t}^{D} (x)|^2  = 1 - G_u
\end{equation}
\begin{equation}
G_{ud} = \sum_x\psi_{t}^{U}(x)  \left( \psi_{t}^{D}(x)  \right)^{*}.
\end{equation}

In Fig.~\ref{fig:se-t-c-r}(a-b), we verify the level of entanglement between the internal and external degrees of freedom during the time evolution of the bipartite system. As initial condition, we establish the quantum walk starts from a state with minimum entropy $S_{\rm e}=0$ corresponding to a separable state. For the clean system, dashed horizontal line, the asymptotic entanglement entropy is $S_{\rm e} \rightarrow 0.872\ldots $ as obtained in Refs.~\cite{carneiro2005entanglement,abal2006quantum}.
That value is quickly overcome for negatively correlated arrangements in $\theta_t$ with  $C=\pm 0.8$. Asymptotically, both cases leads to the maximum value. This is a remarkable result for such a strong correlation, $C=\pm 0.8$. The mathematical proof of that results~\cite{vieira2013dynamically} requires randomness in $\theta_t$ for maximum  $S_e$ as $t \rightarrow \infty$. In the present case, our results show that randomness in the presence of correlation also leads to $S_e  \rightarrow 1$. 

\begin{figure*}[!hbtp]
\centering
\includegraphics[scale=0.49]{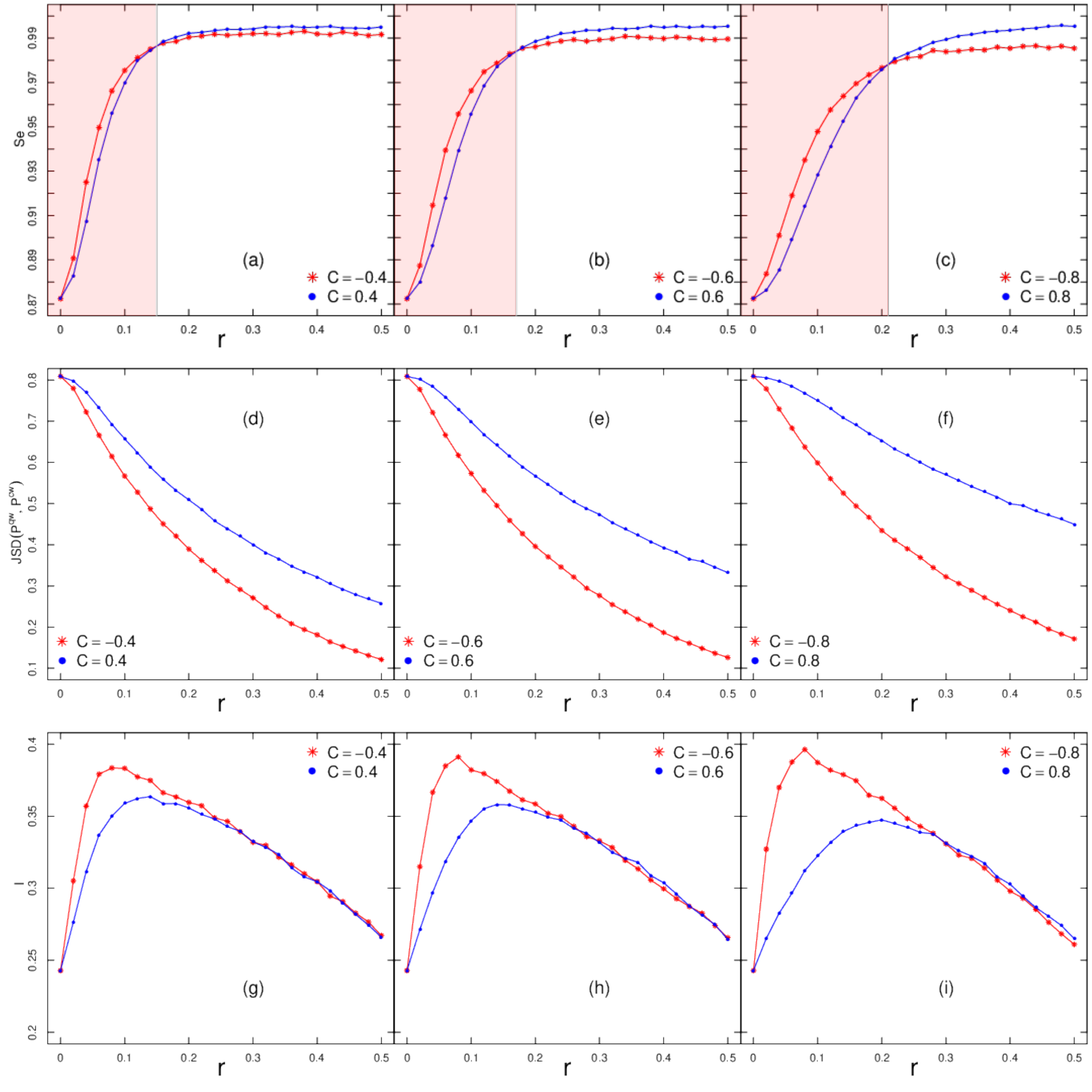}

\caption{Entanglement entropy $S_e$(a-c), Jensen-Shannon dissimilarity JSD (d-f),  Interference measure $I$ (g-i) at $t=500$ for increasing disorder strength $r$ and $C=\{\pm 0.4,\pm 0.6, \pm 0.8\}$.  Each point comes from an average over 500 samples. The  red shadow area for (a-c) shows the regime where  anticorrelation  gives a clear advantage in terms of $S_e$.}
\label{fig:jsd-qc-I-vs-r}
\end{figure*}

Still in Fig.~\ref{fig:se-t-c-r}(a-b), the case with negative correlation $C=-0.8$ produces more entanglement per unit of time  than the case with $C=0.8$.
This novel result persists if the strength of the temporal disorder increases from $r=0.0.5$ ton $r=0.1$. 
How robust is such result for other correlation $C$?
This question is addressed in Fig.~\ref{fig:se-t-c-r}(c-d) where we see that negative correlations are more prone to produce entanglement for any $0<|C|<1$. However, as the intensity of the disorder $r$ increases this advantage fades.
This is further stressed by the results depicted in Fig.\ref{fig:jsd-qc-I-vs-r}(a-c).

Arenas where negative correlations can play a positive role, are rare. Recently, it was discovered configurations in which energy transport at microscopic level can be heightened by negative correlations~\cite{uchiyama2018environmental}. While in our scheme of disorder any value of correlation $|C|<1$ reduces spreading to the diffusive regime, we present a scenario where anti-correlated disorder confers an advantage for entanglement production in comparison with the corresponding protocol with positive correlation.

\begin{figure*}[t]
\centering
\includegraphics[scale=0.59]{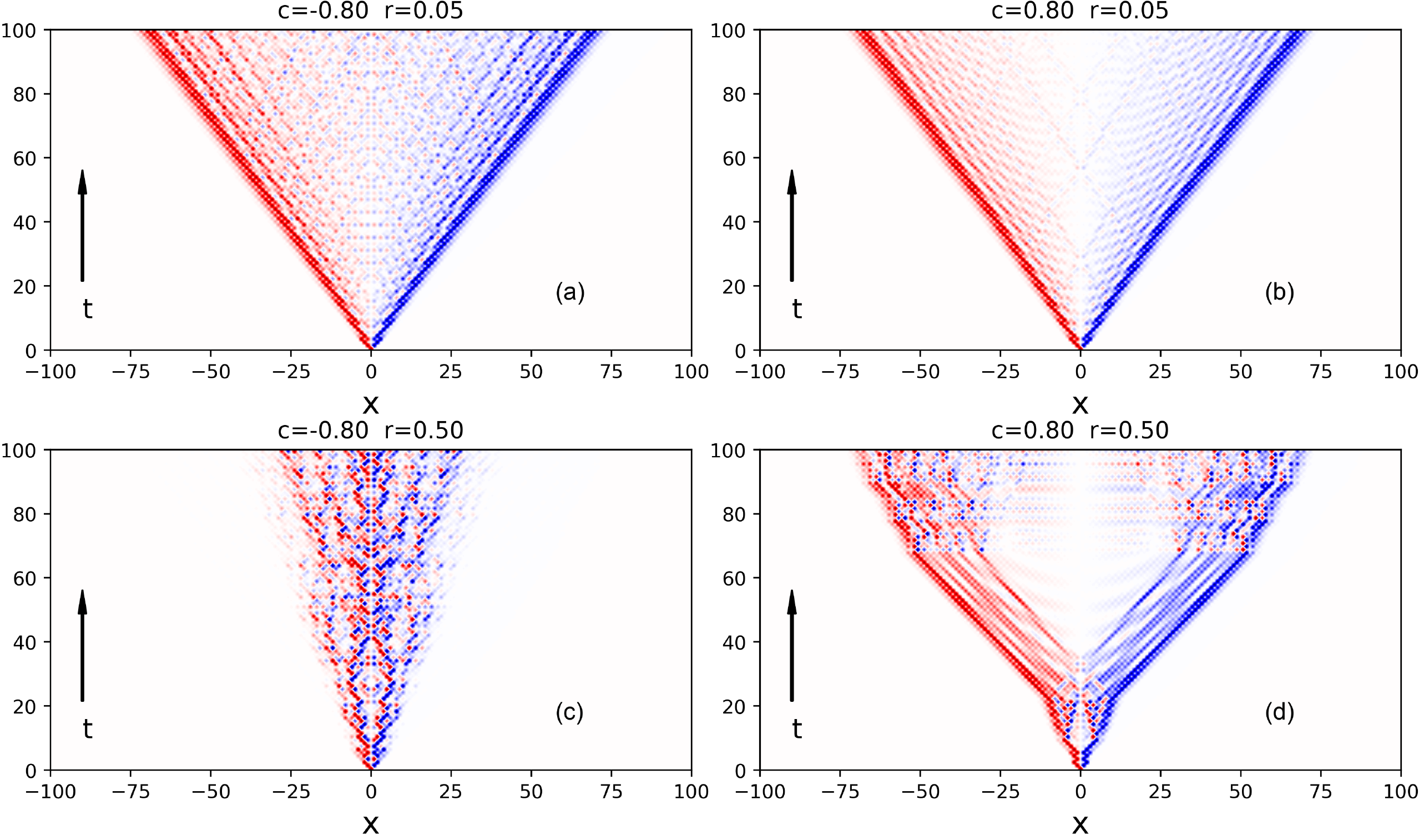}

\caption{Spatiotemporal pattern for the normalized measure
 $A_t(x)/|A_t|^{\max}$  for  $r=\{0.05,0.5\}$  and correlation $C=\pm 0.8$. Red profile:  preponderance of $|\psi_{t}^{D}(x)|$. Blue profile: dominance of $|\psi_{t}^{U}(x)|$. In both cases, the darkness indicates the strength of $A_t(x)/|A_t|^{\max}$. }
\label{fig:mxt45}
\end{figure*}

As aforementioned,  we observed a standard diffusion behavior, Fig.\ref{Fig:alpha-vs-cor}, like a classical system.
Therefore, let us shed light on how much discrepancy is produced between the distributions arising from quantum and classical walks,  $P^{{\rm qw}}_{t}(x)$ and $P^{{\rm cw}}_{t}(x)$. To that, we compute the Jensen-Shannon dissimilarity.
\begin{equation}
JSD_t(P^{{\rm qw}},P^{{\rm cw}}) \equiv 
S(P^{mean})
-S_{mean},
\end{equation}
where the first term is the entropy of the  mean distribution
\begin{equation}
P^{mean}(x) = 
\frac{ P^{{\rm qw}}(x) + P^{{\rm cw}}(x) }{2},
\end{equation}
and $S_{mean}$ is the mean entropy 
\begin{equation}
S_{mean}=
\frac{ S(P^{{\rm qw}}) + S(P^{{\rm cw}}) }
{2}.
\end{equation}
An additional advantage of $JSD_t(P^{{\rm qw}},P^{{\rm cw}})$  refers to its  property of being upper and lower bounded, $0\leq JSD_t(P^{{\rm qw}},P^{{\rm cw}}) \leq 1$.

In Fig.\ref{fig:jsd-qc-I-vs-r}, we see how $JSD_t(P^{{\rm qw}},P^{{\rm cw}})$ changes with  $r$ for typical values of $C$. We generate the space-time probability distribution of the CW using the recursive relation,
\begin{equation}
P^{{\rm cw}}_{t+1}(x)=0.5P^{{\rm cw}}_{t}(x-1) + 0.5P^{{\rm cw}}_{t}(x+1).
\end{equation}
The disorder-free scenario, $r=0$, leads to the highest difference between $P^{{\rm qw}}_{t}(x)$ and $P^{{\rm cw}}_{t}(x)$. 
That dissimilarity decreases as the intensity of the disorder increases, as a result of the increased overlap between the distributions arising from classical and quantum walk.
Clearly, the design of disorder  with negatively correlated patterns in $\theta_t$
is more prone to disturb $P^{{\rm qw}}_{t}(x)$.

Besides $P_t(x)$, we can  delve into the analysis of the relationship between the components $\psi_{t}^{D}(x)$ and $\psi_{t}^{D}(x)$. Allowing for Refs.~\cite{romanelli2004quantum,shikano2014discrete,singh2017interference},we compute $P_{t}^{U}(x)=|\psi_{t}^{U}(x)|^2$
and $P_{t}^{D}(x)=|\psi_{t}^{D}(x)|^2$ in terms of the coin parameter
\begin{align}
P_{t+1}(x) =   P_{t+1}^{U}(x) + P_{t+1}^{D}(x) =
\\
\cos^2\theta_t
\left(  P_{t}^{U}(x+1) + P_{t}^{D}(x-1) \right) 
+ \\
\sin^2\theta_t
\left( P_{t}^{D}(x+1) +  P_{t}^{U}(x-1) \right)    
+ J_{t}(x)
\end{align}
where the local interference term~\cite{singh2017interference} is 
\begin{eqnarray}
J_t(x) & = & \sin2\theta_t 
\Re
\{ \psi_{t}^{U}(x-1) \psi_{t}^{D*}(x-1)  \\
&& -\psi_{t}^{U}(x+1) \psi_{t}^{D*}(x+1),
\}
\end{eqnarray}
where $\Re(z)$ stands for the real part of a complex number $z$ and ${}^*$ its conjugate. The total interference over the chain at $t$ is 
\begin{equation}
I_t = 
\sum_x | J_t(x)   | 
\end{equation}

In Fig.\ref{fig:jsd-qc-I-vs-r}, we see that $I_t$ displays a nonmonotonic behavior, differently from JSD. More importantly,  $I_t$ contains a signature of the regime where the negative correlation overcomes the positive correlation in terms of the entanglement production.
Thus, the emergence of the regime where $S_e(-|C|)>S_e(|C|)$ comes at the expense of the emergence of a marked difference in the mutual 
modulation between the spin components at the level %
$\psi_{t}^{U}(x)$ and $\psi_{t}^{D}(x)$.

In order to better grasp the underlying mechanism behind our results lets inspect the local features of the spatial flux of probability over the chain~\cite{souza2013coin}. This task can be achieved with 
\begin{equation}
A_t(x) =  |\psi_{t}^{U}(x)|^2 - |\psi_{t}^{D}(x)|^2 ,
\label{eq:assymetry}
\end{equation}
In Fig.\ref{fig:mxt45}, we see that the embedded disorder in the coin operation leaves clear fingerprints in the spatiotemporal patterns in the normalized measure $A_t(x)/|A_t|^{\max}$ where $|A_t|^{\max}  = \max_{x} A_t $ is the maximum over the chain updated for each  $t$. For $C=-0.8$, the presence of pulse trains of short-time duration in $\theta_t$ induces a strengthening in the spatial interference of the  components $|\psi_{t}^{U,D}(x)|^2$ which leads to a substantial flux of probability towards the central region. For $C=0.8$, the presence of pulse trains of long-time duration in $\theta_t$ promotes a slowdown in the decay of the peaks near the edges. In both cases $C=\pm 0.8$, the increase in the magnitude of the disorder $r$ leads to a increase in the  evanescence of the fronts near the borders.

\section{\label{sec:remarks}Final remarks}

Entanglement in quantum walks has been addressed  earlier with the design of inhomogeneity either in the coin operator 
\cite{chandrashekar2012disorder,vieira2013dynamically,salimi2012asymptotic,rohde2013quantum,vieira2014entangling,di2016discrete,orthey2019weak,singh2019accelerated,montero2016classical,vieira2014entangling,zeng2017discrete,buarque2019aperiodic,wang2018dynamic} or in the step operator~\cite{sen2019scaling,mukhopadhyay2020persistent,pires2020quantum}. Nevertheless, none of such efforts have found the emergence of the results we have introduced: (i) concerning the time evolution, for weak disorder strength $r<<1$, the production of entanglement entropy per unit of step is higher for negative correlation $C=-c$ than for the opposite case $C=c$; (ii) concerning the asymptotic limit, randomness even in the presence of strong correlation also induces   $S_e  \rightarrow 1$. 
These results open up a new  possibilities for tailoring correlated disorder with target features, while keeping preserved $S_e  \rightarrow 1$.

From an experimental point of view, the optical apparatuses in Ref.~\cite{xue2015experimental,geraldi2019experimental} are promising candidates for implementing our proposal due to their  flexibility in the design of the coin operator.
Of particular interest here is the recent experiment conducted in ~\cite{wang2018dynamic} where their compact photonic platform employs a time-dependent binary disorder in the coin operator that can be adapted to  introduce our prescription of correlation. Their setup allows the reconstruction of the local spinor state for each site which in turn provides an indirect way to quantify the entanglement entropy and related measures.
In a broader view, our work brings about further
prospect in the design of correlated disorder in 
experimental setups for realizing quantum walk \cite{wang2013physical,grafe2016integrated,neves2018photonic}.

We reckon these results contribute to a better understanding of the relationship between correlated disorder and entanglement between the degrees of freedom of a quantum walk. On the one hand, Markov chains are very flexible and diverse with entire books devoted to its features.  On the other hand, quantum walks are multigoal and versatile platforms. By bringing the two fields, the present work strongly suggests that negatively correlated disorder could play a much more important role in applied issues \cite{ambainis2003quantum,portugal2013quantum,venegas2008quantum} as well as fundamental topics\cite{kitagawa2012topological,wu2019topological}.

\begin{acknowledgments}
We acknowledge  financial support from the Brazilian funding agencies CAPES (MAP) as well as CNPq and FAPERJ (SMDQ). 
\end{acknowledgments}

\bibliography{apssamp}

\end{document}